\documentclass[a4paper,fleqn,usenatbib]{mnras}

\usepackage{newtxtext,newtxmath}

\usepackage[T1]{fontenc}
\usepackage{ae,aecompl}

\usepackage{graphicx}	
\usepackage{amsmath}	
\usepackage{amssymb,url}	

\def\simlt{\ \raise -2.truept\hbox{\rlap{\hbox{$\sim$}}\raise5.truept   %
\hbox{$<$}\ }}
\def\simgt{\ \raise -2.truept\hbox{\rlap{\hbox{$\sim$}}\raise5.truept   %
\hbox{$>$}\ }}                                                          %

\newcommand{\pd}[3]{\frac{\partial^{#3} #1}{\partial {#2}^{#3}}} 
\newcommand{\td}[3]{\frac{d^{#3} #1}{d {#2}^{#3}}} 
\renewcommand{\v}[1]{\ensuremath{\mathbf{#1}}} 
\newcommand{\gv}[1]{\ensuremath{\mbox{\boldmath$ #1 $}}}
\renewcommand{\bar}[1]{\ensuremath{\overline{#1}}}

\title[]{Through A Mini Halo, Darkly}

\author[G. Beck and S. Colafrancesco]{
Geoff Beck,$^{1}$\thanks{E-mail: geoffrey.beck@wits.ac.za}
Sergio Colafrancesco,$^{1}$\thanks{In honour of the memory of my colleague, who passed before this work was complete}
\\
$^{1}$School of Physics, University of the Witwatersrand, 1 Jan Smuts Avenue, 2001, South Africa\\
}

\date{Accepted XXX. Received YYY; in original form ZZZ}

\pubyear{2017}

\begin{document}
\label{firstpage}
\pagerange{\pageref{firstpage}--\pageref{lastpage}}
\maketitle

\begin{abstract}
We explore the effects of the scattering of photons incident on a dark matter halo through their interaction with either electrons or photons produced by dark matter annihilation. Particularly, we examine the effects of this scattering upon the observed spectrum of a distant AGN or of the Cosmic Microwave Background. Our results indicate that ultra-compact mini halos and other dense dark matter clumps can induce an observable Comptonisation of AGN spectra as well as a Sunyaev-Zel'dovich effect (SZE) with an optical depth similar to that attained by thermal electrons in the Coma cluster. The rate of occlusion of the line of sight to a distant AGN by these dense mini-halos is also estimated using various limits existing on the population of dark compact bodies.  
\end{abstract}




\section{Introduction}

Previous work on the Comptonisation effects of the CMB resulting from electrons produced non-thermally by Dark Matter (DM) annihilation, in particular the DM-induced Sunyaev-Zel'dovich Effect (SZE), had concluded that it may be a promising approach to DM hunting, but, the expected signals would be fainter than local thermal electron Comptonisation in most cases~\citep{sz-dm-colafrancesco,colafrancesco-dm-sze1,colafrancesco-dm-sze2,sz-dm-lavalle}. It has been pointed out, however, that small compact structure like dwarf galaxies may be ideal targets for observation of DM-induced Comptonisation~\citep{colafrancesco-dm-sze1} as they lack strong baryonic emission and are relatively compact objects; the amplitude of the effects is very sensitive however to assumptions about the DM halo profile~\citep{sz-dm-lavalle}.

In this work we propose that Ultra-Compact Mini Halos (UCMHs) make even more attractive targets for hunting a DM-induced Comptonisation effect of the CMB  as their extreme central density allows for a far greater optical depth.

UCMHs are highly interesting structures, as they form very early in universe~\citep{ricotti2009,bringmann2012} and have been shown to potentially provide constraints on the many different physical processes/phenomena that may give rise to them. These include cosmic strings and other topological defects~\citep{silk1993}, as well as small-scale effects of inflation~\citep{aslanyan2015}, or even phase-transitions in the early universe~\citep{josan2010}. The large central density of UCMHs ensures that they can persist easily into the present epoch~\citep{berezinsky2006,berezinsky2008,bringmann2012}. However, such objects are difficult to detect, as they have low mass, their DM-induced emissions are rapidly made faint even for distances well inside the Milky-Way halo, and they are expected to contain very little baryonic matter~\citep{ricotti2009,bertschinger1985}. Therefore, it is desirable to find additional methods for probing the nature of such objects due to their potential as probes of small-scale cosmology and inflation.

We demonstrate here that such compact structures would have potentially observable impacts on the CMB through the production of an SZE as well as being able to Comptonize the spectrum of a distant AGN (out to $z \sim 1$) if they were to impinge upon the line of sight to it. 
We also calculate a conservative rate of encounters between such AGNs and UCMHs. Furthermore, these effects potentially provide an additional avenue to put constraints on UCMHs which are otherwise so difficult to probe directly by limiting the number density of such objects according to SZE observations as well as potentially limiting the fraction of DM in UCMHs via null-observations of the form of AGN variation discussed.

This work is structured as follows: in section~\ref{sec:methods} we discuss the spectral models, the Comptonization processes, the UCMH characteristics, and DM annihilation process. In sections~\ref{sec:results} and \ref{sec:results2} we show the results of our computations, and we will present our conclusions in section~\ref{sec:conc}.

\vspace{-0.5cm}

\section{Methods and simulations}
\label{sec:methods}
We simulate the passage of a beam of photons emitted from a distant source passing through an ultra-compact mini-halo. We will detail below the various aspects of these calculations.

\subsection{AGN Spectra}
\label{sec:agn}

The spectrum employed for the X-ray emission of an AGN is a power-law with a high-E cut-off: 
\begin{equation}
S(E) = A E^{-\Gamma} \exp{\left(-\frac{E}{E_c}\right)}\;.
\end{equation}
The relevant parameters are taken from \citet{agn1} assuming $\Gamma = 1.7$, $80 \leq E_c \leq 160$ keV and that the luminosity in the 17-60 keV range is $10^{41} \leq L_x \leq 10^{43}$ erg s$^{-1}$ (which determines $A$ based on the luminosity distance value at redshift $z$). To consider more complete spectra, we make use of spectral shapes from \citet{mrk421} and we normalise the flux appropriately for a given redshift of the source.


\subsection{Ultra-compact Mini Halos}
\label{sec:ucmh} 

We will briefly review here the details of the UCMH model that are pertinent to our study.

UCMHs have been shown to plausibly collapse from over densities as small as $\delta \sim 10^{-3}$ in the redshift range $200 \leq z \leq 1000$~\citep{ricotti2009}, and  this was expanded more rigorously in \citet{bringmann2012}. Their extreme central density allows them to survive any tidal stripping by larger structures~\citep{bringmann2012} and persist into the present epoch for all the masses of UCMH that we will consider here.

Since, after kinetic decoupling of the WIMP from the Standard Model particles (which occurs after the WIMP abundance has frozen out~\citep{bringmann2009}), UCMHs form entirely via radial in-fall~\citep{ricotti2009,ricotti2008,bringmann2009} (as their is no significant scattering of WIMPs by other matter) this leads to a density profile of the form
\begin{equation}
\rho_{UCMH} (r,z) = \frac{3 f_{\chi} M_{UCMH} (z)}{16 \pi R_{UCMH}^{\frac{3}{4}}(z) r^{\frac{9}{4}}} \; , \label{eq:rho}
\end{equation}
where $f_{\chi}$ is the fraction of matter in the form of DM, $M_{UCMH} (z)$ is the UCMH mass, and $R_{UCMH}(z)$ is the effective radius of the UCMH at redshift $z$. 
This effective radius is given by the following expression derived from numerical simulations~\citep{ricotti2007,ricotti2008}
\begin{equation}
R_{UCMH} (z) = 0.019 \left(\frac{1000}{1+z}\right) \left( \frac{M_{UCMH}(z)}{\mbox{M}_{\odot}}\right)^{\frac{1}{3}} \; . 
\label{eq:size}
\end{equation}
The extremely steep UCMH density profile $\sim r^{-9/4}$ is a consequence of the spherically symmetric collapse, as derived in \citet{fillmore1984,bertschinger1985} and later demonstrated by numerical simulations~\citep{vogelsberger2009,ludlow2010}. It must be noted that $M_{UCMH}$ is assumed fixed for $z \leq 10$ following \citet{bringmann2012}, so for all the values of $z$ we study here $M_{UCMH} (z) = M_{UCMH} (0)$. 

Recent results~\citep{ucmhrho1,ucmhrho2} have argued that UCMHs cannot attain the $\rho \propto r^{-9/4}$ density profile in realistic N-body simulations of halo formation. Therefore, we will also follow \citet{ucmhrho2} in using the following profile as a comparison to the $\rho \propto r^{-9/4}$ case
\begin{equation}
\rho_{UCMH} = \frac{\rho_s}{(r/r_s)^{3/2}(1 + r/r_s)^{3/2}} \; ,
\end{equation}
where $\rho_s$ and $r_s$ are found following \citet{ucmhrho2}.

It is clear that any such a profile cannot be valid all the way down to $r = 0$. Thus, some minimum radius must be established within which the density flattens and becomes constant. This is due to the fact that accretion at later times in the halo's life will largely influence the outer regions of the UCMH~\citep{ricotti2009,bertschinger1985}.

Thus, the DM density within the UCMH will be limited by
\begin{equation}
\rho (r \leq r_{min}) = \rho(r_{min}) \; ,
\end{equation}
where $r_{min}$ is defined by the expression from \citet{bringmann2012} as
\begin{equation}
r_{min} = 2.9\times 10^{-7} \left(\frac{1000}{1 + z_c}\right)^{2.43} \left( \frac{M_{UCMH}(0)}{\mbox{M}_{\odot}}\right)^{-0.06} R_{UCMH}(0) \; ,
\end{equation}
where $z_c$ is the redshift at which the UCMH collapsed.
The effects of WIMP annihilation will also limit the cuspiness of the UCMH, with the maximum possible density being taken as
\begin{equation}
\rho_{max} = \frac{m_{\chi}}{\langle \sigma V\rangle (t-t_i)} \; .
\label{eq:rhomax}
\end{equation}
This limit on the density defines a cut-off radius $r_{cut}$ such that
\begin{equation}
\rho (r \leq r_{cut}) = \rho_{max} \, 
\label{eq:rhomax2}
\end{equation}
where $m_{\chi}$ is the mass of the WIMP with a thermally-averaged annihilation cross section of $\langle \sigma V\rangle$.

For UCMHs at the present epoch, $t$ is 13.799 Gyr~\citep{planck2014} and $t_i$ is taken to be equal to $t_{eq}$ following the arguments given in \citet{scott2009,wright2006}. Thus, if $r_{cut} < r_{min}$ the cut-off radius and maximal density is determined by $r_{min}$. But, if $r_{cut} > r_{min}$, then the core density of the UCMH will be limited by Eq.(\ref{eq:rhomax2}), and will automatically satisfy the $r_{min}$ requirement as well. 

Then, given a fraction of DM found in the form of UCMHs $f$, we can find the rate of occlusion events in an environment with DM density $\rho_{DM}$ in analogy to micro-lensing
\begin{equation}
\Gamma_{UCMH} = 2 f \bar{R}_{UCMH} v_{UCMH} \int_l dl \frac{\rho_{DM}}{M_{UCMH}}  \; , \label{eq:tau}
\end{equation}
where $l$ is the line of sight through the host galaxy, $v_{UCMH}$ is the UCMH orbital speed, and $\bar{R}_{UCMH}$ is the radius at which $\tau_c$ is 10\% of its maximum value in order to capture only the densest part of the halo where the Comptonisation will be strongest. 

\subsection{Dark Matter Annihilation}
The source function for particle $i$ (electrons/positrons or photons) with energy $E$ from a DM annihilation is taken to be
\begin{equation}
Q_i (r,E) = \langle \sigma V\rangle \sum\limits_{f}^{} \td{N^f_i}{E}{} b_f \left(\frac{\rho_{\chi}(r)}{m_{\chi}}\right)^2 \; ,
\end{equation}
where $r$ is distance from the halo centre, $\langle \sigma V \rangle$ is the non-relativistic velocity-averaged annihilation cross-section, $f$ labels the annihilation channel intermediate state with a branching fraction $b_f$ and differential $i$-particle yield $\td{N^f_i}{E}{}$, $\rho_{\chi}(r)$ is the DM radial density profile, and $m_{\chi}$ is the WIMP mass. The $f$ channel used as a test case will be annihilation via quarks $q\bar{q}$.

The yield functions $\td{N^f_i}{E}{}$ are taken from \citet{ppdmcb1,ppdmcb2} (with electro-weak corrections).

This source function will be used to calculate photon distribution via
\begin{equation}
N_{\gamma,\chi} (E) = \int_0^{R_{UCMH}} dr \, 4 \pi r^2 Q_{\gamma} (r,E) t_{cross}\; ,
\end{equation}
with $R_{UCMH}$ being the UCMH radius and $t_{cross}$ being the halo crossing time. The electron distribution is defined as 
\begin{equation}
N_{e,\chi} (E) = \left(\td{n_{e^-}}{E}{} + \td{n_{e^+}}{E}{}\right) \; ,
\end{equation}
where $\td{n_{e^\pm}}{E}{}$ are found via the stationary solution to the equation
\begin{equation}
\begin{aligned}
\pd{}{t}{}\td{n_e}{E}{} = & \; \gv{\nabla} \left( D(E,\v{r})\gv{\nabla}\td{n_e}{E}{}\right) + \pd{}{E}{}\left( b(E,\v{r}) \td{n_e}{E}{}\right) + Q_e(E,\v{r}) \; ,
\end{aligned}
\end{equation}
where $D(E,\v{r})$ is the diffusion coefficient, $b(E,\v{r})$ is the energy loss function, and $Q_e(E,\v{r})$ is the electron source function from DM annihilation/decay. In this case, we will work under the simplifying assumption that $D$ and $b$ lack a spatial dependence and thus we will include only average values for magnetic field and thermal electron densities. For details of the solution see \citet{Colafrancesco2007}.\\
We thus define the functions as follows~\citep{Colafrancesco1998}
\begin{equation}
D(E) = \frac{1}{3}c r_L (E) \frac{\overline{B}^2}{\int^{\infty}_{k_L} dk P(k)} \; ,
\end{equation}
where $\overline{B}$ is the average magnetic field, $r_L$ is the Larmour radius of a relativistic particle with energy $E$ and charge $e$ and $k_L = \frac{1}{r_L}$. This, combined with the requirement that
\begin{equation}
\int^{\infty}_{k_0} dk P(k) = \overline{B}^2 \; ,
\end{equation}
where $k_0 = \frac{1}{d_0}$, with $d_0$ being the smallest scale on which the magnetic field is homogeneous, yields the final form
\begin{equation}
D(E) = D_0 d_0^{\frac{2}{3}} \left(\frac{\overline{B}}{1 \mu\mbox{G}}\right)^{-\frac{1}{3}} \left(\frac{E}{1 \mbox{GeV}}\right)^{\frac{1}{3}}  \; , \label{eq:diff}
\end{equation}
where $D_0 = 3.1\times 10^{28}$ cm$^2$ s$^{-1}$, and we assume that $d_0 = 1 pc$ for the UCMHs we consider.

The energy loss function is defined by
\begin{equation}
\begin{aligned}
b(E) = & b_{IC} E^2 (1+z)^4 + b_{sync} E^2 \overline{B}^2 \\&\; + b_{Coul} \overline{n} (1+z)^3 \left(1 + \frac{1}{75}\log\left(\frac{\gamma}{\overline{n} (1+z)^3}\right)\right) \\&\; + b_{brem} \overline{n} (1+z)^3 \left( \log\left(\frac{\gamma}{\overline{n} (1+z)^3 }\right) + 0.36 \right) \;,
\end{aligned}
\label{eq:loss}
\end{equation}
where $\gamma$ is the Lorentz factor of an electron, $\overline{n}$ is the average thermal electron density in the UCMH and is given in cm$^{-3}$, while $b_{IC}$, $b_{synch}$, $b_{col}$, and $b_{brem}$ are the inverse Compton, synchrotron, Coulomb and bremsstrahlung energy loss factors, taken to be $0.25$, $0.0254$, $6.13$, and $1.51$ respectively in units of $10^{-16}$ GeV s$^{-1}$~\citep{Colafrancesco2007}. Here $E$ is the energy in GeV and the B-field is in $\mu$G. We will assume $ \overline{B} = 1$ $\mu$G, and $\overline{n} = 10^{-6}$ cm$^{-3}$ for the UCMHs we consider. These values are chosen in keeping with the low electron density of dwarf galaxies and the average fields observed in larger structures~\citep{mwbfield}. We do this because UCMHs are not expected to contain much baryonic matter and so the magnetic field will be that of the host environment and the plasma density will be taken to be similar to the value found in dwarf galaxies in similar host environments (to account for some environmental baryonic material). Lower baryon densities will increase the Compton optical depth for electrons produced by DM annihilation as the baryon density contributes to energy-loss for the DM-produced electrons.
All of the calculations in this work will feature diffusion, as it has a very significant impact in such small structures as UCMHs. Neglecting diffusion results in the optical depth from DM-produced electrons being over-estimated by a factor $\mathcal{O}(10^2)$.

\subsection{Spectral effects}
\label{sec:spec}

The two spectral modifications that we will examine will be produced by absorption of AGN emitted photons by pair-production with gamma-ray photons produced by DM annihilation, and by Comptonization of the both the CMB background and AGN photons off of DM-produced electrons/positrons.

For pair production between incident photons and those produced by DM annihilation we calculate the optical depth $\tau_{\gamma\gamma}$ following \citet{dwek2012} and use $S_{\gamma\gamma} (E_{\gamma}) = S_0(E_{\gamma}) \exp{(-\tau_{\gamma\gamma})}$ where $S_0$ is the intrinsic spectrum of the distant AGN.
\begin{equation}
\epsilon_{th} = \frac{2 \left(m_e c^2\right)^2}{E_{\gamma} (1-\mu)} \; ,
\end{equation}
where $\mu = \cos{\theta}$, $\theta$ is the angle between the interacting photons, and $E_\gamma$ is incoming photon energy.

The cross-section for photon absorption by pair production is given by~\citep{dwek2012}
\begin{equation}
\sigma_{\gamma\gamma} (E_{\gamma}, \epsilon, \mu) = \frac{3\sigma_T}{16}(1-\beta^2)\left[ 2\beta (\beta^2-2) + (3-\beta^4)\ln{\frac{1+\beta}{1-\beta}}\right] \; ,
\end{equation}
where $\sigma_T$ is the Thompson cross-section, $\epsilon$ is energy of the DM-produced photon, and
\begin{equation}
\beta = \sqrt{1 - \frac{\epsilon_{th}}{\epsilon}} \; .
\end{equation}

The optical depth for incoming photons due to pair production is then~\citep{dwek2012}
\begin{equation}
\tau_{\gamma \gamma} (E_{\gamma}) = \int_{-1}^{1} d\mu \, \frac{1-\mu}{2} \int_{\epsilon_{th}}^{\infty} d\epsilon \, \frac{N_{\gamma,\chi} (\epsilon)}{4\pi R_{UCMH}^2} \sigma_{\gamma\gamma} (E_{\gamma},\epsilon,\mu) \; .
\end{equation}

The resulting spectrum of a distant source modified by pair-production with DM-produced gamma-ray photons is then given by
\begin{equation}
S_{\gamma\gamma} (E_{\gamma}) = S_0(E_{\gamma}) \exp{(-\tau_{\gamma\gamma})} \; ,
\end{equation}
where $S_0$ is the intrinsic spectrum of the distant AGN.

For the case of the Comptonisation (and the SZE) by DM-produced electrons the optical depth is found via
\begin{equation}
\tau_{C} (E_{\gamma}) = \sigma_T \int d l  \int dE \, N_{e,\chi} (E) \; .
\end{equation}
Thus, to first-order in $\tau_{C}$ the shifted spectrum is given by 
\begin{equation}
S_{C} (E_{\gamma}) = (1 - \tau_{C}) S_0(E_{\gamma}) + \tau_{C} \int ds \, P_1(s) S_0(E_{\gamma} \mbox{e}^{-s}) \; ,
\end{equation}
where $s = \ln{\left(\frac{E^{\prime}}{E_{\gamma}}\right)}$ is the logarithmic energy shift of the incoming photon, and the single scattering redistribution function $P_1 (s)$ is given by
\begin{equation}
P_1 (s) = \int_0^\infty d p \, f_e(p) P_s(s,p) \; ,
\end{equation}
where $p = \frac{v}{c}\gamma$ is the dimensionless electron momentum, $f_e (p)$ is the normalised electron momentum distribution with $\int_0^\infty d p \, f_e(p) = 1$, and $P_s (s,p)$ is probability of an electron with momentum $p$ inducing a logarithmic shift in the photon energy of $s$ (see \citet{ensslin2000} for details). The effects of this Comptonisation will be explicitly non-thermal~\citep{colafrancesco2003}.


\section{Results - SZE}
\label{sec:results}

We start showing the results for the test case of DM particles with mass $100$ GeV 
annihilating via quarks with $\langle \sigma V\rangle = 3 \times 10^{-26}$ cm$^{3}$ s$^{-1}$ 
producing an SZE  through the DM-electrons produced by annihilations in a mini-halo situated along the line of sight to the observer.
\begin{figure}
\centering
\resizebox{0.9\hsize}{!}{\includegraphics[scale=0.7]{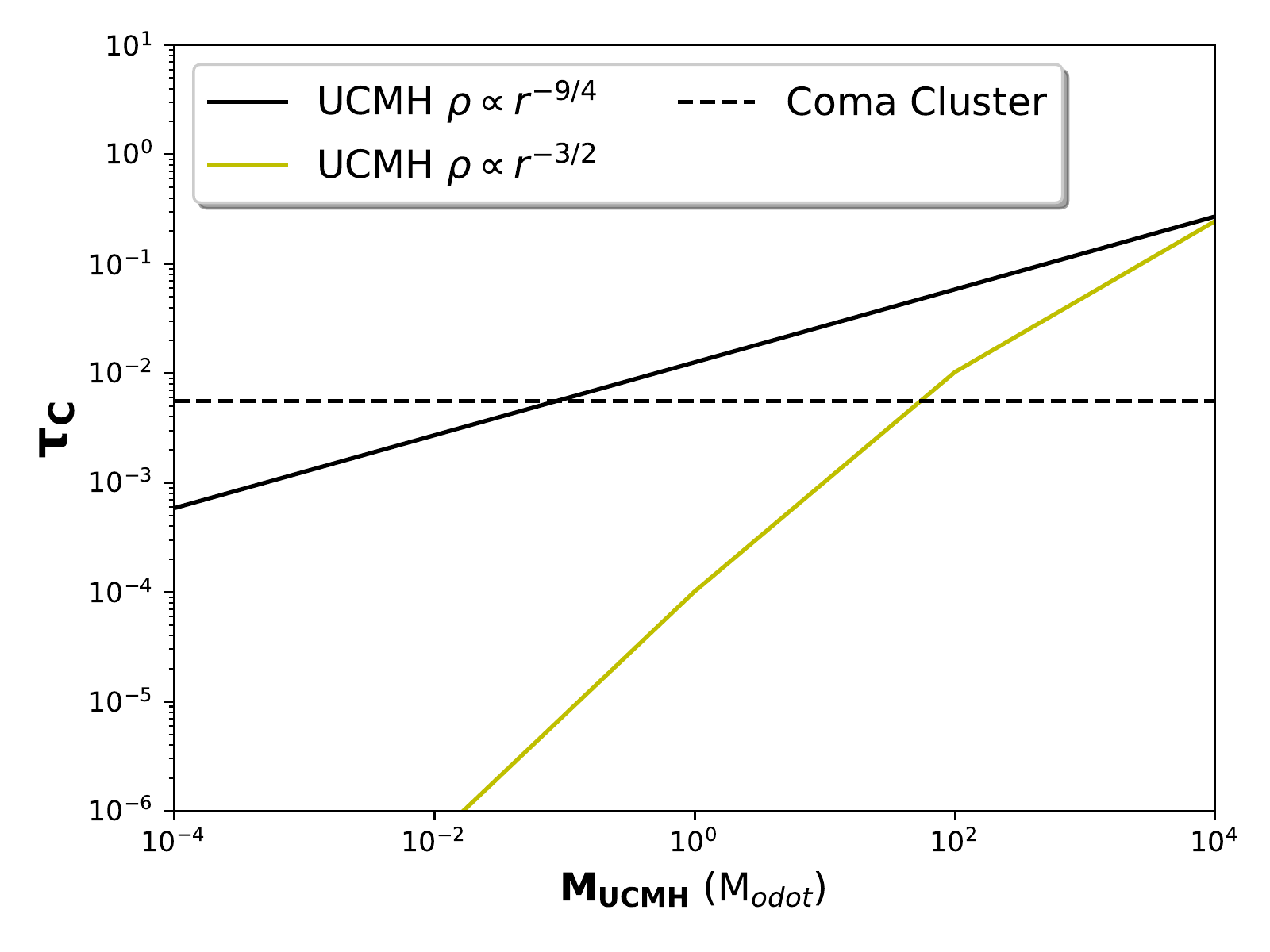}}
\caption{The DM-induced optical depth as a function of the UCMH mass, compared to that found for the a thermal electron distribution in the Coma cluster as in \citet{comatau}.}
\label{fig:tau}
\end{figure}

In Figure~\ref{fig:tau} we see that the Compton optical depth of DM-produced electrons in a UCMH scales as $\tau_C \propto M_{UCMH}^{1/3}$. This means that $\tau_C$ reaches a value similar to the thermal electron optical depth in the Coma cluster for  $M_{UCMH} \sim 10^{-2}$ M$_{\odot}$, and begins to exceed this value for $M_{UCMH} \ge 10^{-1}$ M$_{\odot}$. Despite the tiny size of the UCMH, such structure is capable of generating a similar optical depth from DM annihilations as found in a galaxy cluster $\approx 10^{14}$ times heavier. This is due to the extreme value of the UCMH central density  which can be illustrated as follows for $M_{UCMH} = 10^4$ M$_{\odot}$: the core radius $r_{cut} \sim 4 \times 10^5$ $r_{Earth}$ with $\rho_{max} \sim 2 \times 10^{10}$ GeV cm$^{-3}$. For smaller mass UCMHs $\rho_{max}$ is constant but $r_{cut} \propto M_{UCMH}^{1/3}$ so that the minimum size for $r_{cut}$ to obtain values of $\tau_C$ similar to the Coma cluster is $\sim 4.1\times 10^3$ $r_{Earth}$ (for $M_{UCMH} = 10^{-2}$ M$_{\odot}$). For the $\rho \propto r^{-9/4}$ density profile, the central core radius acts to set the maximum injection rate of electrons from DM annihilation, as $\rho_{max}$ is independent of the halo size for the studied mass range with this density profile. It must be noted that, unlike in previously studied halos~\citep{colafrancesco-dm-sze1,colafrancesco-dm-sze2,sz-dm-lavalle}, the form of Eq.~(\ref{eq:rhomax}) means that $\tau_{C}$ will be amplified for heavier WIMPs, but only for the very compact $\rho \propto r^{-9/4}$ density profile and for $\rho \propto r^{-3/2}$ when $M_{UCMH} \gtrsim 10^2$ M$_{\odot}$. This means that there is the potential for a significant Comptonization from UCMH objects for a range of halo and WIMP masses. We note that the effect of using the $\rho \propto r^{-3/2}$ density profile depends strongly on the UCMH mass. In this case the optical depth only passes that of Coma for objects larger than $10$ M$_{\odot}$. However, $1$ M$_{\odot}$ UCMHs already attain $\tau_C \sim 10^{-4}$ whereas a similar calculation for a non-thermal electron population produced via  DM annihilation in the Coma cluster yields only $\tau_C \sim 10^{-7}$.

In order to detect an SZE signal in UCMHs we need to resolve such DM halos adequately. At $z \sim 1$ this requires milli-arcsecond resolution; however, within $\sim 10$ Mpc of Earth we can resolve halos above $1 M_{\odot}$ at arcsecond levels \big(the angular size scales as $M_{UCMH}^{1/3}$\big) which should be achievable for a large range of masses with both ALMA~\footnote[1]{\url{http://almascience.nrao.edu/documents-and-tools/cycle5/alma-technical-handbook/view}} and Millimetron~\citep{millimetron}, with the potential to reach milli-arcsecond resolution via sub-millimeter VLBI~\cite{millimetron}.

The flux amplitude of the SZE negative peak, for $\rho \propto r^{-9/4}$, integrated over the angular area of the UCMH, at $\sim 150$ GHz is $3\times 10^{-4}$ of the CMB peak amplitude if $M_{UCMH} = 10^{-4}$ M$_{\odot}$, with the positive part of the SZE at $\sim 500$ GHz being an order of magnitude smaller. On the other end of our mass scale, for a UCMH with $10^4$ M$_{\odot}$ mass, the negative SZE peak reaches $0.1$ of the CMB maximum amplitude. In the case of the $\rho \propto r^{-3/2}$ profile the amplitude of the SZE follows a pattern relative to $\rho \propto r^{-9/4}$ that is very similar to the relation between the optical depths for the two density profiles from Fig.~\ref{fig:tau}. This implies that the major limiting factor to observation is resolving the effect spatially, as the magnitude of the effect can be substantial. A local population can be estimated with upper limits from reionisation~\citep{reion-ucmh} while assuming a mono-chromatic mass-distribution of UCMHs, this order of magnitude estimate indicates as many as $10^{14}$ $10^{-4}$ M$_{\odot}$ or $10^6$ $10^4$ M$_{\odot}$ halos within just the Milky-Way. Using gamma-ray limits from \citet{bringmann2012} (which apply for a 1 TeV WIMP with $\langle \sigma V \rangle = 3\times 10^{-26}$ cm$^3$ s$^{-1}$) we find that there are at most $5\times 10^{12}$ $10^{-2}$ M$_{\odot}$ halos or $40$ with a mass of $10^4$ M$_{\odot}$. An important factor limiting observation would be the problem of source confusion and cosmological backgrounds. However, investigating this would require a detailed analysis that is currently beyond the scope of this work.

\section{Results - AGN}
\label{sec:results2}

Here we show the results for the test case of DM particles with mass $100$ GeV 
annihilating via quarks with $\langle \sigma V\rangle = 3 \times 10^{-26}$ cm$^{3}$ s$^{-1}$ 
producing a spectral distortion of an AGN spectrum via a mini-halo situated along the line of sight to the observer.

We consider two effects: Comptonization of the AGN spectrum produced by DM-produced electrons in the UCMH, and pair production produced by AGN emitted photons interacting with the gamma-ray photons produced in the DM annihilation.
In contrast to the Compton optical depth from DM annihilation discussed previously, that for pair-production between incoming photons from an AGN spectrum studied here and DM-produced gamma-rays is negligible, meaning that no halo can provide a spectral absorption effect (even for photons experiencing Compton scattering prior to interaction with DM photons). 
Therefore, only the Comptonization results will be shown for the AGN spectrum.

\begin{figure}
\centering
\resizebox{0.9\hsize}{!}{\includegraphics[scale=0.7]{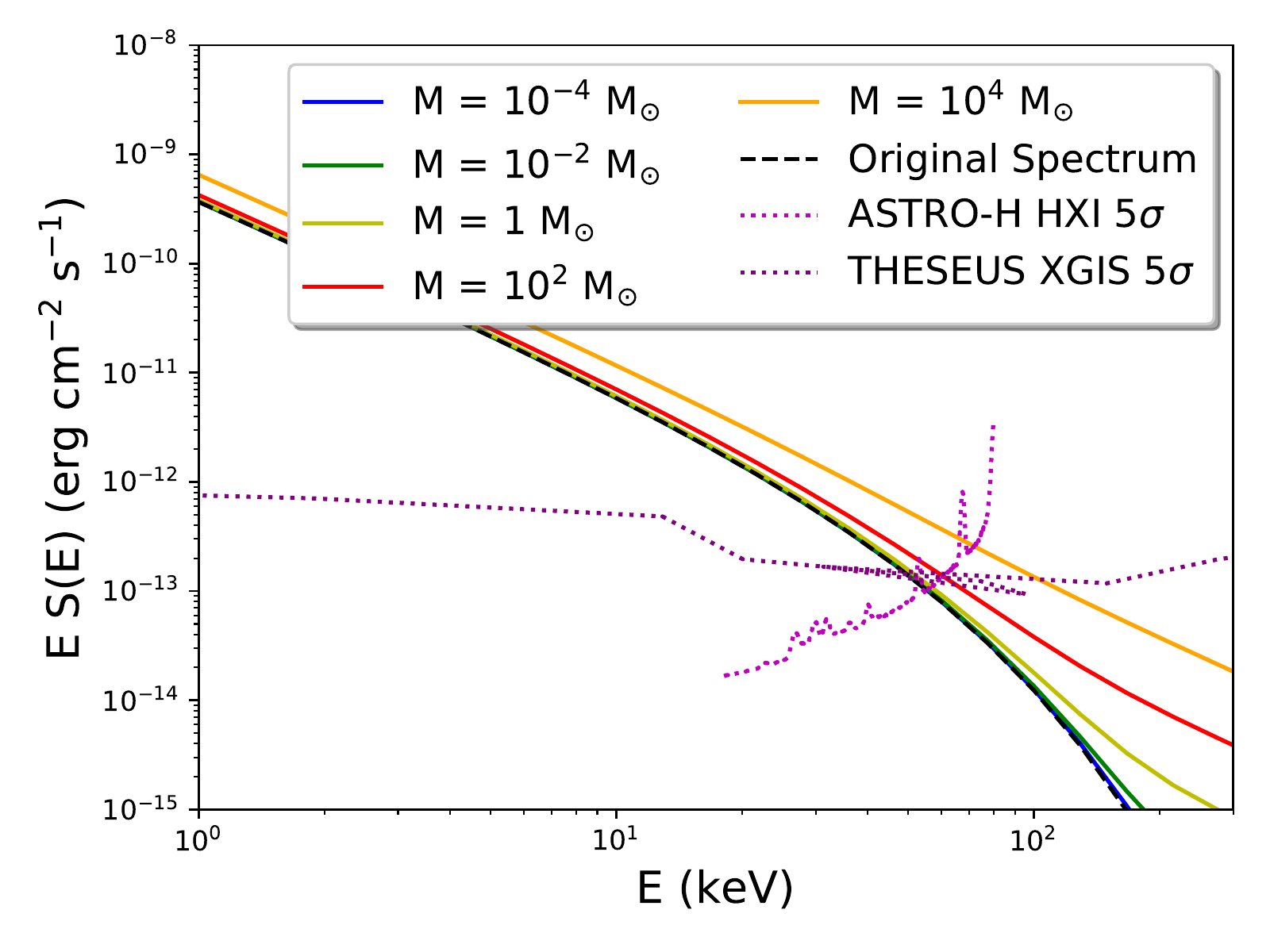}}
\caption{Spectral effects of Comptonisation by UCMHs (with the $\rho \propto r^{-9/4}$ profile) of various mass compared to X-ray telescopes sensitivity. This is calculated for an AGN and UCMH at $z\sim 1$ with a 17 - 60 keV luminosity of $10^{43}$ erg s$^{-1}$, and a cut-off energy at 80 keV. Also shown are detection limits for THESEUS~\citep{theseus} and ASTRO-H\protect\footnotemark[2], both for 1 Ms exposure time.}
\label{fig:spectrum}
\end{figure}\footnotetext[2]{\url{http://astro-h.isas.jaxa.jp/researchers/sim/sensitivity.html}}


In Figure~\ref{fig:spectrum} we show the modifications to the high-E tail of an AGN spectrum caused by the passage through UCMHs of various mass where we assume that both the halo and the AGN are at $z\sim 1$, the AGN has a 17-60 keV luminosity of $10^{43}$ erg s$^{-1}$, and a cut-off energy at 80 keV. The DM annihilation-produced population of electrons was determined assuming $\langle \sigma V\rangle = 3 \times 10^{-26}$ cm$^3$ s$^{-1}$. 
As long as the UCMH mass remains above $1$ M$_{\odot}$ ($10$ M$_{\odot}$ when $\rho \propto r^{-3/2}$), then there is an extension of the power-law spectrum beyond the intrinsic cut-off, visible within the energy band of a project like ASTRO-H or THESEUS (ASTRO-H being defunct is used merely as a benchmark for future X-ray experiments). This visibility would be in the form of a temporary fluctuation within the minimum-flux observation window of the telescopes, during the UCMH occlusion, from the original spectrum displayed in the figure. Notably, for the smaller halos, the extension is in the fashion of a broken power-law, whereas larger halo masses simply continue the power-law spectrum unbroken.

Using a complete spectrum similar to that from \citet{mrk421} we also studied the modification of the rest of the AGN spectrum as a consistency check. Unlike the above case, the low energy cut-off becomes more severe (i.e. moves to higher energies) as low energy photons are lost to Comptonisation. As expected, however, this is at least an order of magnitude less apparent than the modification to the high-energy cut-off shown in Fig.~\ref{fig:spectrum}. At the inflection point between synchrotron and the synchrotron-self-Compton spectra the flux is boosted by a factor that varies between 1.004 and 1.4 over the mass range $10^{-4}$ to $10^4$ M$_{\odot}$. Thus, the most prominent effect is to be found in the high-energy cut-off, due to the large difference in photon energies between the extremities of the spectrum.

%


A necessary question is how common the occlusion of an AGN by UCMH actually is. The rate of UCMH encounters per AGN and per year is estimated using Eq~(\ref{eq:tau}) 
for the case of a galaxy with an NFW DM halo density profile that has characteristic scale radius of 20 kpc, a characteristic density of $10^4$ times the critical density of the universe (virial mass $1.3 \times 10^{11}$ M$_{\odot}$ and virial radius $0.112$ Mpc), and an average orbital velocity within it of 200 km s$^{-1}$. These values are chosen to produce an `unremarkable' (in the sense that it does not exhibit extreme properties that might exaggerate the rate of AGN occlusions) galaxy with similar properties to the Milky-Way. This yields the following rates per AGN per year
\begin{equation}
\Gamma_{UCMH} = \begin{cases}
\sim 10^{-8} M_{UCMH}^{-2/3} f \; \mbox{yr$^{-1}$} & \rho_{UCMH} \propto r^{-9/4} \\
\sim 10^{-7} M_{UCMH}^{-2/3} f \; \mbox{yr$^{-1}$} & \rho_{UCMH} \propto r^{-3/2} 
\end{cases} \; .
\end{equation}
These results suggests that such events are around three orders of magnitude rarer in occurrence than estimates for binary neutron star mergers per galaxy~\citep{chrulinska2017} for $\rho_{UCMH} \propto r^{-3/2}$ with reionisation limits~\citep{reion-ucmh}. With the rate being slightly smaller for the more intense Comptonisation of the $\rho_{UCMH} \propto r^{-9/4}$ density profile. If the limits from \citep{bringmann2012} from local gamma-ray searches are considered then a more complex dependence on $M_{UCMH}$ is introduced, reducing the rate by a factor of $10^{2}$ for the lightest halos considered and by $10^4$ for the largest halos. The rate of these events is strongly dependent on $f$ and may serve as a means of constraining $f$ outside the local environment. Actual constraints would need to be derived from a detailed analysis of a large sample of AGN spectral histories if $f$ is expected to be small (as suggested by existing results~\cite{reion-ucmh,bringmann2012}). We note however that both \cite{reion-ucmh,bringmann2012} assume the $r^{-9/4}$ profile and would be weakened by the use of the $r^{-3/2}$ case, whereas the constraints from $\Gamma_{UCMH}$ above grow stronger in latter case.

\section{Conclusions}
\label{sec:conc}

We have shown that there are potentially observable consequences both of a UCMH impinging upon the line of sight to a distant source, like an AGN, as well as an induced SZE upon the CMB spectrum that is resolvable at the $\sim$ arcsecond level within $\sim 10$ Mpc of Earth. 
Notably, neither of these effects are as significant in conventional DM halos, because the resulting optical depth is far too small. A further important point is that the probable lack of baryonic matter in a UCMH (due to tidal stripping) means that these effects will not be significantly obscured by purely baryonic emissions within the UCMH structure.
The SZE in particular can be used to limit UCMH abundances in the local environment, as an expected number density resolvable at arcsecond levels, an achievable resolution for sub-millimetre observatories such as ALMA or Millimetron, can be determined for a given UCMH fraction $f$. This is of course subject to a caveat regarding the effects of source and background confusion that are not studied here. Unlike these Comptonization effects, no significant pair-production process can be induced by photons from DM annihilation.

The occlusion of the line of sight to a distant AGN by such a compact body is shown to have potentially dramatic consequences producing potentially large modifications to spectral cut-offs during the period of occlusion. These events were shown to be considerably rarer than binary neutron star mergers per galaxy. However, because this effect is potentially accessible out to redshifts $z \sim \mathcal{O}(1)$, a large volume can be studied for the characteristic Comptonisation of the AGN spectrum. Null results for this variation of AGN spectra could also be used to limit the UCMH fraction $f$ outside the local environment.

The use of the shallower density profile $\rho_{UCMH} \propto r^{-3/2}$ that has been suggested in the literature has a strong effect only for smaller mass UCMH objects and so does not substantially alter the arguments made with the $\rho_{UCMH} \propto r^{-9/4}$ profile, as the studied effects were most prominent for larger UCMH objects, above $1$ M$_{\odot}$.

Finally, these effects yield a promising new approach to the search for such compact DM halos, which is usually complicated by their small size, lack of baryonic matter, and faint fluxes from potential annihilation/decay at cosmological distances.

\section*{Acknowledgements}
This work is based on the research supported by the South African
Research Chairs Initiative of the Department of Science and Technology
and National Research Foundation of South Africa (Grant No 77948). 




\bibliographystyle{mnras}
\bibliography{ucmh} 




\bsp	
\label{lastpage}
\end{document}